%% file: sample-sigplan.tex
\documentclass[sigplan,screen]{acmart}
\usepackage{bbm}
\usepackage{ulem}
\usepackage{amsmath}
\usepackage[flushleft]{threeparttable}
\usepackage{multirow} 
\usepackage{enumitem}
\DeclareMathOperator*{\argmax}{arg\,max}

\usepackage{lipsum}
\newcommand\blfootnote[1]{%
  \begingroup
  \renewcommand\thefootnote{}\footnote{#1}%
  \addtocounter{footnote}{-1}%
  \endgroup
}

\expandafter\def\expandafter\normalsize\expandafter{%
    \normalsize%
    \setlength\abovedisplayskip{-4pt}%
    \setlength\belowdisplayskip{4pt}%
    \setlength\abovedisplayshortskip{-8pt}%
    \setlength\belowdisplayshortskip{2pt}%
}

\AtBeginDocument{%
  \providecommand\BibTeX{{%
    \normalfont B\kern-0.5em{\scshape i\kern-0.25em b}\kern-0.8em\TeX}}}

\setcopyright{acmlicensed}
\copyrightyear{2024}
\acmYear{2024}
\acmDOI{XXXXXXX.XXXXXXX}

\acmConference[DAC'24]{61st ACM/IEEE Design Automation Conference}{June 23--27, 2024}{San Francisco, CA}

\acmSubmissionID{1122}

\copyrightyear{2024}
\acmYear{2024}
\setcopyright{acmlicensed}\acmConference[DAC '24]{61st ACM/IEEE Design
Automation Conference}{June 23--27, 2024}{San Francisco, CA, USA}
\acmBooktitle{61st ACM/IEEE Design Automation Conference (DAC '24), June
23--27, 2024, San Francisco, CA, USA}
\acmDOI{10.1145/3649329.3658482}
\acmISBN{979-8-4007-0601-1/24/06}

\begin{document}


\newcommand{\name}{3D-Carbon}

\title{ {\name}: An Analytical Carbon Modeling Tool for 3D and 2.5D Integrated Circuits}

\author{\small Yujie Zhao$^1$, Yang (Katie) Zhao$^{2*}$, Cheng Wan$^1$, Yingyan (Celine) Lin$^1$}
\affiliation{%
  \institution{$^1$\textit{Georgia Institute of Technology}, $^2$\textit{University of Minnesota, Twin Cities}}
  \city{\{eiclab.gatech, chwan, celine.lin\}@gatech.edu, yangzhao@umn.edu}
  \country{}
}




\input{Sections/0-Abstract}



\maketitle

\blfootnote{$^*$ Corresponding author.}
\renewcommand{\shortauthors}{Yujie Zhao, et al.}
\input{Sections/introducation}

\input{Sections/background}

\input{Sections/model}
\input{Sections/validation}
\input{Sections/conclusion}

\vspace{-0.5em}
\bibliographystyle{ACM-Reference-Format}
\bibliography{sample-base}

\end{document}

%% file: Sections/0-Abstract.tex
\begin{abstract}
Environmental sustainability is crucial for Integrated Circuits (ICs) across their lifecycle, particularly in manufacturing and use. Meanwhile, ICs using 3D/2.5D integration technologies have emerged as promising solutions to meet the growing demands for computational power. However, there is a distinct lack of carbon modeling tools for 3D/2.5D ICs. Addressing this, we propose 3D-Carbon, an analytical carbon modeling tool designed to quantify the carbon emissions of 3D/2.5D ICs throughout their life cycle. 3D-Carbon factors in both potential savings and overheads from advanced integration technologies, considering practical deployment constraints like bandwidth. We validate 3D-Carbon's accuracy against established baselines and illustrate its utility through case studies in autonomous vehicles. We believe that 3D-Carbon lays the initial foundation for future innovations in developing environmentally sustainable 3D/2.5D ICs. Our open-source code is available at \url{https://github.com/UMN-ZhaoLab/3D-Carbon}.

\end{abstract}
\vspace{-12em}


%% file: Sections/introducation.tex
\vspace{-1.8em}
\section{Introduction}
\label{sec:intro}
\begin{figure}[t]
  \centering
  \includegraphics[width=1\linewidth]{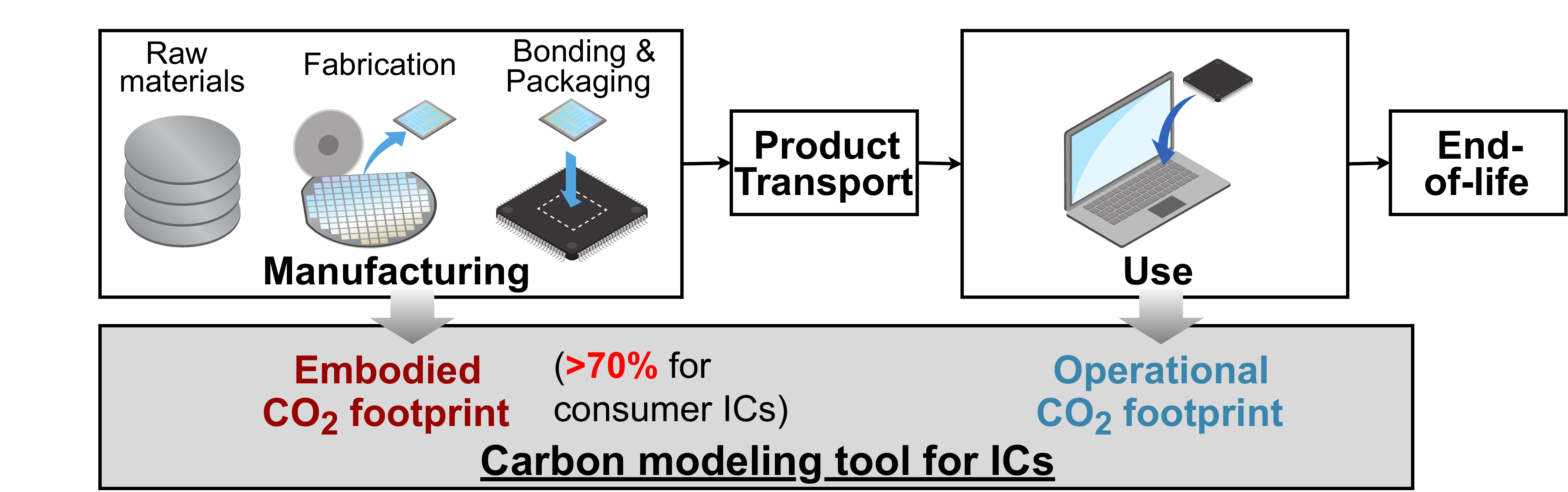}
\vspace{-2.2em}
\caption{A carbon modeling tool tracks embodied and operational carbon emissions throughout ICs' lifecycle~\cite{gupta2022act}.}
\vspace{-1.2em}
\label{fig:life_cyc}
\end{figure}

For decades, Moore's Law has driven Integrated Circuits (ICs), enhancing their computational power, reducing size and cost, and improving energy efficiency. These advancements have been crucial for developing new technologies like artificial intelligence (AI). However, the carbon emissions from ICs, covering their entire lifecycle from manufacturing to disposal (see Fig.~\ref{fig:life_cyc}), pose significant environmental sustainability challenges. Recent reports indicate that the carbon of Information and Communication Technology represents 2.1\%$\sim$3.9\% of global greenhouse gas emissions~\cite{freitag2021real}.

To ensure environmental sustainability in ICs, addressing their carbon emissions throughout their lifecycle, particularly in \textit{manufacturing} and \textit{use} phases, is crucial~\cite{gupta2022act}. Currently, the \textit{embodied carbon} from manufacturing activities often surpasses the \textit{operational carbon} from energy consumption during the use of today's ICs~\cite{gupta2022act}. 
This shift is due to extensive operational energy efficiency improvements over years. Embodied carbon can represent over 70\% of the total carbon footprint for consumer ICs~\cite{gupta2022act,elgamal2023design}.

In parallel, as the pace of Moore’s Law in 2D monolithic ICs has slowed in recent years, significant advancements have been made in developing 3D/2.5D ICs to meet rising computational demands. These advanced ICs offer notable improvements over 2D ICs in power efficiency, performance, and area in various commercial products~\cite{feng2022chiplet,kim2021micro,w2w,3Dvcache}.

However, currently, there's no comprehensive tool to evaluate the carbon footprint of 3D/2.5D ICs, particularly their embodied carbon, which is crucial due to their growing complexity: While additional manufacturing steps increase carbon emissions per wafer, factors like improved yield, area efficiency, use of heterogeneous technologies, and fewer metal layers could reduce the overall carbon footprint.

Current tools for estimating the embodied carbon of 3D/2.5D ICs have limitations. Industry-based databases rely on data about materials~\cite{epyclca} or manufacturing processes~\cite{gupta2022act,elgamal2023design}, but their practicality and accuracy are limited by the availability of up-to-date carbon emission data. First-order approaches, such as the one in~\cite{eeckhout2022first}, estimate the embodied footprint per chip based on die size. Another method, ACT+~\cite{elgamal2023design}, estimates 2.5D IC carbon footprint from 2D ICs based on cost comparison and simplistically treats 3D stacked dies as 2D. These techniques provide general insights but lack the detailed breakdowns needed for effective carbon-conscious design.

\begin{figure*}[h]
  \centering
  \includegraphics[width=1\linewidth]{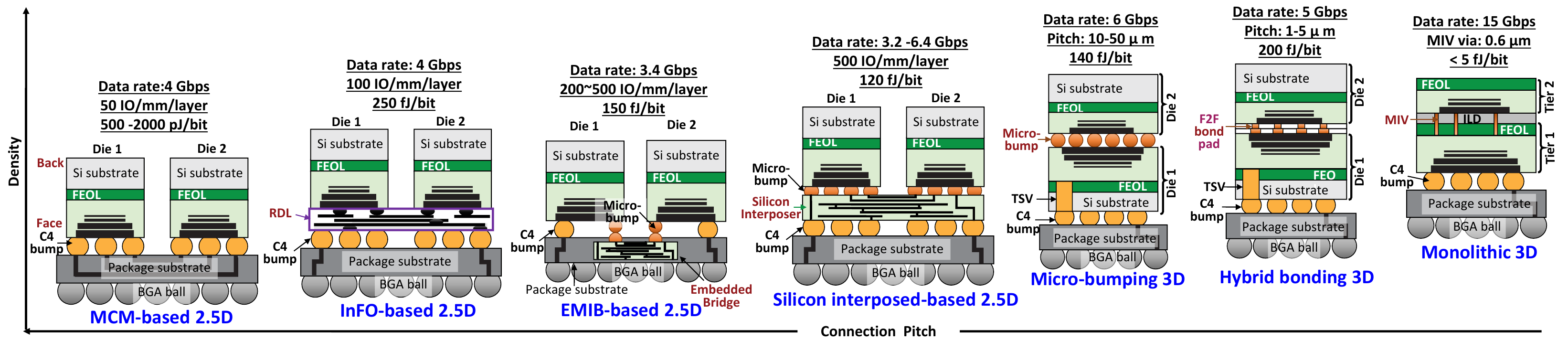}
 \vspace{-2.5em}
\caption{The vertical stack diagram of 3D and 2.5D integration options studied in this paper.}
\label{fig:3D_vertical}
 \vspace{-1.em}
\end{figure*}
 
To comprehensively quantify the overall carbon footprint of 3D/2.5D ICs, we propose {\name}, an analytical carbon modeling tool, in this paper. Our contributions are as follows: 

\begin{itemize}[leftmargin=*,topsep=0mm]

\item We develop {\name}, an analytical carbon modeling tool for various 3D/2.5D ICs. We believe that {\name} lays the initial foundation for future innovations in developing environmentally sustainable 3D/2.5D ICs.

\item Using {\name}, we can predict the embodied carbon emissions, including the overhead associated with advanced integration technologies, and estimate the operational carbon emissions during use through surveyed parameters or third-party operational energy estimation plug-ins.

\item We evaluate {\name} and demonstrate its valuable insights and broad applicability through case studies in autonomous vehicles to guide sustainable decision-making in choosing or replacing ICs.
\end{itemize}

%% file: Sections/background.tex
\section{Background}
\label{sec:background}

\subsection{Commercial 3D/2.5D Integration Technologies}
\label{sec:background_intro_3D}

\begin{table}[t]
\centering
\caption{3D/2.5D integration technologies summary.}
\vspace{-1em}
\begin{threeparttable}

\centering

\resizebox{1\linewidth}{!}{
\begin{tabular}{|c|c|c|c|c|c|c|}
\hline
    \multirow{3}{*}{\begin{tabular}[c]{@{}c@{}}\textbf{2.5}\\\textbf{or}\\\textbf{3D}\end{tabular}} & \multirow{3}{*}{\begin{tabular}[c]{@{}c@{}}\textbf{Integration}\\\textbf{Technology}\end{tabular}} & \multirow{3}{*}{\begin{tabular}[c]{@{}c@{}}\textbf{F2F}\\\textbf{or}\\\textbf{F2B}\end{tabular}} & \multirow{3}{*}{\textbf{Stacking}} & \multirow{3}{*}{\begin{tabular}[c]{@{}c@{}}\textbf{\# of}\\\textbf{max 3D}\\\textbf{dies/tiers}\end{tabular}} & \multirow{3}{*}{\begin{tabular}[c]{@{}c@{}}\textbf{Representative}\\\textbf{Manufacturers}\end{tabular}} & \multirow{3}{*}{\begin{tabular}[c]{@{}c@{}}\textbf{Representation}\\\textbf{Products}/\\\textbf{Prototypes}\end{tabular}} \\
    & & & & & &  \\ 
    & & & & & &  \\ 
    \hline \hline
    \multirow{9}{*}{3D} & \multirow{4}{*}{\begin{tabular}[c]{@{}c@{}}Hybrid\\Bonding\end{tabular}} & \multirow{2}{*}{F2F} & \multirow{2}{*}{D2W/W2W} & \multirow{2}{*}{2} & \multirow{4}{*}{\begin{tabular}[c]{@{}c@{}}TSMC's\\SoIC-X~\cite{tsmc_23_3dfabric}\\Intel's\\Fevores Direct~\cite{mahajan2021advanced}\end{tabular}} & \multirow{2}{*}{\begin{tabular}[c]{@{}c@{}}AMD\\ 3D V-cache~\cite{3Dvcache}\end{tabular}} \\
    & & & & & & \\ \cline{3-5}\cline{7-7}
    & & \multirow{2}{*}{F2B} & \multirow{2}{*}{D2W/W2W} & \multirow{2}{*}{$\geq$2} &  &   \multirow{2}{*}{\begin{tabular}[c]{@{}c@{}}AMD Ryzen\\7-5800X3D~\cite{3Dvcache}\end{tabular}}\\ 
    & & & & & & \\ \cline{2-7}
    
    & \multirow{4}{*}{\begin{tabular}[l]{@{}l@{}}Micro-\\bumping\end{tabular}} & \multirow{2}{*}{F2F} &  \multirow{2}{*}{D2W/W2W} &  \multirow{2}{*}{2}   & \multirow{4}{*}{\begin{tabular}[c]{@{}c@{}}TSMC's\\SoIC-P~\cite{tsmc_23_3dfabric}\\Intel's\\Fevores~\cite{mahajan2021advanced}\end{tabular}} & \multirow{2}{*}{\begin{tabular}[c]{@{}c@{}}Intel Lakefield\\Core i5-L16G7~\cite{lakefield}\end{tabular}} \\
    & & & & & & \\ \cline{3-5}\cline{7-7}
    & & \multirow{2}{*}{F2B} & \multirow{2}{*}{D2W/W2W} & \multirow{2}{*}{$\geq$2} &    & \multirow{2}{*}{\begin{tabular}[c]{@{}c@{}}HBM~\cite{w2w}\end{tabular}}\\ 
    & & & & & & \\ \cline{2-7}
    
    & \multirow{1}{*}{M3D} & \multirow{1}{*}{F2B} & \multirow{1}{*}{N/A} & \multirow{1}{*}{2} & \multirow{1}{*}{} & \multirow{1}{*}{RISC-V Core~\cite{kim2021micro}} \\ \hline \hline
   
    \multirow{6}{*}{2.5D} & \multirow{2}{*}{\begin{tabular}[c]{@{}c@{}}MCM\end{tabular}}  & \multirow{2}{*}{N/A} & \multirow{2}{*}{N/A} & \multirow{2}{*}{N/A} & \multirow{2}{*}{\begin{tabular}[c]{@{}c@{}}AMD's Infinity\\Fabric~\cite{w2w}\end{tabular}}  & \multirow{2}{*}{\begin{tabular}[c]{@{}c@{}}AMD EPYC\\7000 series~\cite{w2w}\end{tabular}} \\ 
    & & & & & & \\ \cline{2-7} 
    
    & \multirow{2}{*}{InFO} & \multirow{2}{*}{N/A} & \multirow{2}{*}{N/A} & \multirow{2}{*}{N/A} & \multirow{2}{*}{\begin{tabular}[c]{@{}c@{}}TSMC's InFO-2.5D\\CoWoS-L/R~\cite{tsmc_23_3dfabric}\end{tabular}}  & \multirow{2}{*}{\begin{tabular}[c]{@{}c@{}}AMD\\ Ravi 31~\cite{gpu_navi31}
    \end{tabular}}\\ 
    & & & & & & \\  \cline{2-7}  
    & \multirow{2}{*}{\begin{tabular}[c]{@{}c@{}}EMIB\end{tabular}}  & \multirow{2}{*}{N/A} & \multirow{2}{*}{N/A} & \multirow{2}{*}{N/A} & \multirow{2}{*}{\begin{tabular}[c]{@{}c@{}}Intel's \\EMIB~\cite{EMIB}\end{tabular}}  & \multirow{2}{*}{\begin{tabular}[c]{@{}c@{}}Stratix 10~\cite{IntelStratix10_2024} \end{tabular}} \\ 
    & & & & & & \\ \cline{2-7} 
    
    & \multirow{2}{*}{\begin{tabular}[c]{@{}c@{}}Silicon\\Interposer\end{tabular}}  & \multirow{2}{*}{N/A} & \multirow{2}{*}{D2W} & \multirow{2}{*}{N/A} & \multirow{2}{*}{\begin{tabular}[c]{@{}c@{}}TSMC's\\CoWoS-S~\cite{tsmc_23_3dfabric}\end{tabular}}   & \multirow{2}{*}{\begin{tabular}[c]{@{}c@{}}Nvidia GPU\\P100~\cite{gpu_p100}\end{tabular}}\\ 
    & & & & & & \\ \hline
\end{tabular}
}
\end{threeparttable}
\vspace{-1em}
\label{tab:list_of_integrations}
\end{table}

We examine three commercial 3D integration technologies and four 2.5D integration technologies (see Tab.~\ref{tab:list_of_integrations} and Fig.~\ref{fig:3D_vertical}). 

\subsubsection{\underline{\textbf{3D Integration}}}
\textbf{Micro-bumping 3D.} This method stacks multiple dies vertically using micron-level solder balls for 3D connections, with a larger pitch than other 3D technologies.
\textbf{Hybrid Bonding 3D.} This technique stacks two 2D dies using bond pads through the metal layers.
\textbf{Monolithic 3D (M3D).} M3D employs sequential manufacturing to create fine-pitched MIVs (typically $<$0.6$\mu$m) for inter-tier connections~\cite{kim2021micro}.
This paper emphasizes block-level M3D partitioning, where functional blocks like memory and logic macros are separated into different tiers, enabling the use of existing 2D EDA processes~\cite{akgun2019network}.

\subsubsection{\underline{\textbf{2.5D Integration}}} 
\textbf{Multi-chip module (MCM).} This method places multiple pre-designed dies on an organic package substrate.
\textbf{Integrated fan-out (InFO).} InFO, evolving from fan-out wafer-level packaging, uses a redistribution layer (RDL) as the substrate, offering smaller line space than MCM and including chip-first and chip-last approaches.
\textbf{Embedded Multi-die Interconnect Bridge (EMIB).} 
EMIB integrates multiple dies within a single package using a silicon interconnect bridge.
\textbf{Silicon interposer.} Utilizing a silicon substrate (passive or active~\cite{3Dvcache}), silicon interposers provide the finest line space but may increase carbon costs.

\subsection{IC Carbon and Carbon Optimization Metrics}
\subsubsection{\underline{\textbf{IC Total Life Cycle Carbon}}}
Following the industry-endorsed frameworks~\cite{gupta2022act,elgamal2023design}, we estimate an IC's total life cycle carbon footprint as follows, including both operational ($C_{operational}$) and embodied ($C_{emb}$) emissions:

\vspace{-0.7em}
\begin{align}
C_{total} = C_{operational} +C_{emb}
\label{eq:carbon_total}
\end{align}

\vspace{-0.7em}
\subsubsection{\underline{\textbf{Indifference Point and Breakeven Time Analysis}}}


The embodied carbon ($C_{emb}$) of 3D/2.5D ICs includes both potential savings (e.g., fewer back-end-of-line (BEOL) layers) and overheads (e.g., advanced integration manufacturing processes). Meanwhile, their operational carbon ($C_{operational}$) benefits from shorter interconnect lengths but faces higher power needs for interfaces. Thus, 3D/2.5D ICs don't always offer clear sustainability benefits in both $C_{emb}$ and $C_{operational}$.

To aid in sustainable decision-making for choosing and replacing 3D/2.5D ICs over 2D ICs for fixed workload applications, we use the indifference point metric ($T_c$) and the breakeven time metric ($T_r$) as defined in~\cite{greenchip}, which are based on the embodied carbon costs ($C^{2D}_{emb}$/$C^{3D/2.5D}_{emb}$), the use-phase carbon intensity ($CI_{ues}$),  and the power consumptions ($P^{2D}_{app}/{P^{3D/2.5D}_{app}}$, see Eq.~(\ref{eq:power})):

\begin{align}
\hspace{-5pt}T_c=&\frac{C_{emb}^{3D/2.5D}-C_{emb}^{2D}}{CI_{use}(P^{2D}_{app}-P^{3D/2.5D}_{app})},
T_r=\hspace{-2pt}\frac{C_{emb}^{3D/2.5D}}{CI_{use}(P^{2D}_{app}-P^{3D/2.5D}_{app})}
\label{eq: breaktime}
\end{align}

In scenarios of ``choosing'', $T_c$ indicates when the saved $C_{emb}$ of 3D/2.5D ICs is offset by increased $C_{operational}$.  In scenarios of ``replacing'', $T_r$ shows the breakeven time when the increased $C_{emb}$ of 3D/2.5D ICs is compensated by reduced $C_{operational}$ with the assumption that the embodied carbon cost has already been invested for the 2D ICs. We compare $T_c$/$T_r$ to the IC's remaining lifetime ($T_{life}$) to guide sustainable decision-making in choosing or replacing ICs.

%% file: Sections/model.tex
\section{{\name} Modeling Tool}

\subsection{{\name}: High-Level Overview}
\label{sec:carbon_model_overall}

\begin{figure}[t]
  \centering
  \includegraphics[width=0.88\linewidth]{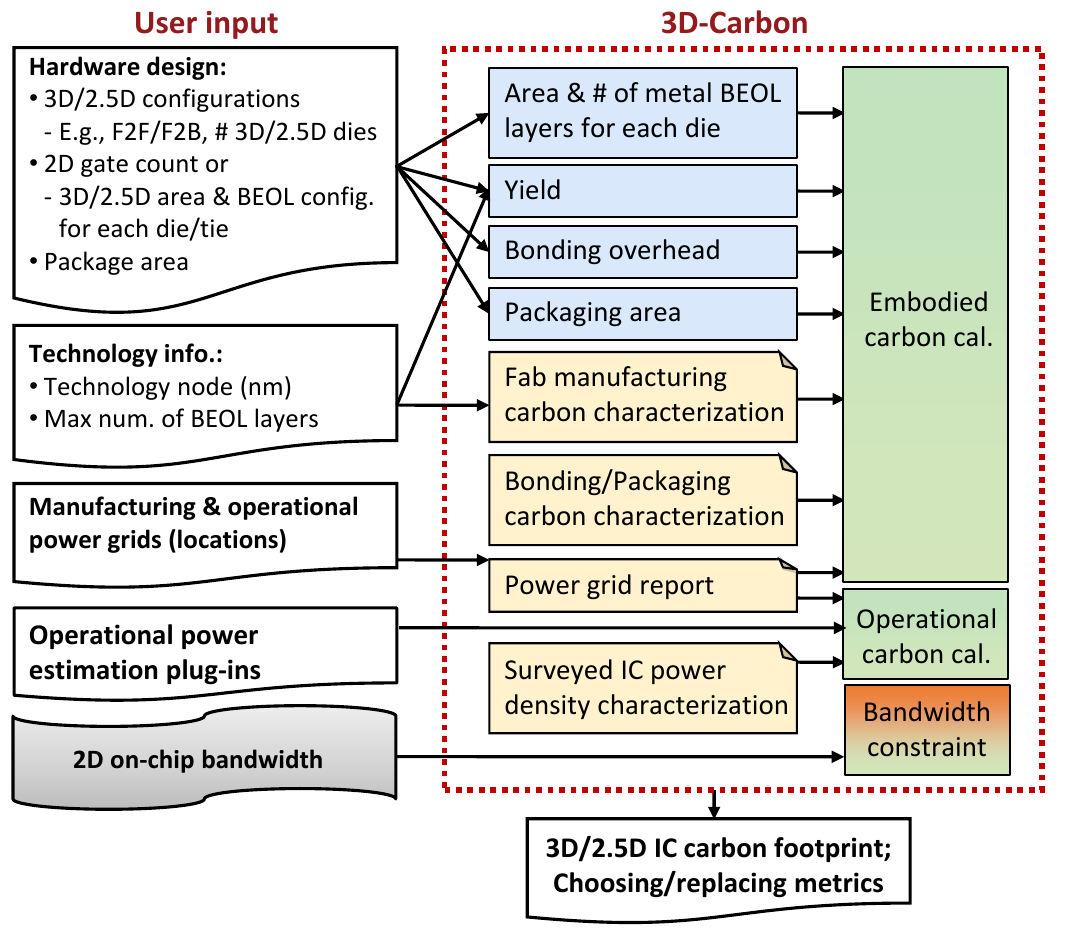}
\vspace{-1.5em}
\caption{The overview of the proposed {\name}.}
\vspace{-1.2em}
\label{fig:2.5D/3Dintro}
\end{figure}

Fig.~\ref{fig:2.5D/3Dintro} shows an overview of {\name}, concentrating on both the embodied and operational carbon emissions of 3D/2.5D ICs. The relevant parameters have been obtained from industry environmental reports, as listed in Tab.~\ref{tab:list_of_parameters}. 

For estimating \textit{embodied carbon emission}, {\name} uses a hardware design description consisting of 3D/2.5D configurations and IC area details, a technology information description of the technology node and the maximum number of BEOL layers, and the manufacturing location.


For calculating \textit{operational carbon emission}, {\name} integrates with operational power consumption plugins like~\cite{guler2020mcpat,gpudensity,cpudensity} and utilizes the use-location's carbon intensity.

Additionally, 3D/2.5D ICs employ off-die I/O interfaces for data movements, which utilize on-chip wire resources in 2D ICs. Thus, {\name} introduces an I/O bandwidth constraint to assess the viability of 3D/2.5D ICs compared to their 2D counterparts in terms of data movements.

\subsection{{\name}: Embodied Carbon Emission}
\label{sec:carbon_model_embodied}

Unlike 2D ICs, 3D/2.5D ICs integrate $N$ 2D dies with additional bonding emissions. For interposer-based ICs, an extra RDL/silicon interposer is manufactured. We calculate the overall embodied emission ($C_{emb}^{3D/2.5D}$) by summing the emissions from die manufacturing ($C_{die}^{3D/2.5D}$), bonding ($C_{bonding}^{3D/2.5D}$), packaging ($C_{packaging}^{3D/2.5D}$), and the 2.5D interposer ($C_{int}^{2.5D}$):

\vspace{-0.2em}
\begin{align}
C_{emb}^{3D/2.5D} = C_{die}^{3D/2.5D} + C_{bonding}^{3D/2.5D} + C_{packaging}^{3D/2.5D}+ C_{int}^{2.5D}  
\label{eq:carbon_embodied_3D/2.5D_overall}
\end{align}

\vspace{-0.5em}
\subsubsection{$C_{die}^{3D/2.5D}$ Model}
\label{wafer}

To calculate the embodied carbon from die manufacturing in 3D/2.5D ICs, we begin from the carbon emission per wafer ($C_{wafer_{die_i}}$) for each die (denoted as $die_i$), the corresponding die-per-wafer count ($DPW_{die_{i}}$), and the yield ($Y_{die_{i}}$) (as detailed in Sec. \ref{sec: yield}):

\vspace{-0.7em}
\begin{align}
\label{eq: carbon_die}
C_{die}^{3D/2.5D} &= \sum_{i=1}^N \frac{C_{wafer_{die_i}}}{DPW_{die_{i}}} \cdot \frac{1}{Y_{die_{i}}}
\end{align}

We calculate the die-per-wafer count ($DPW_{die_{i}}$) by dividing wafer area ($A_{wafer_{die_i}}$) by die area ($A_{die_{i}}$)~\cite{stow2016cost}, similarly applied to interposer-per-wafer count ($DPW_{int}$):

\begin{table}[t]
\centering
\caption{3D/2.5D IC embodied carbon model parameters.}
\begin{threeparttable}
\vspace{-1em}
\resizebox{0.9\linewidth}{!}
{
\begin{tabular}{|c|c|c|}
\hline 
    \textbf{Parameter} & \textbf{Range} & \textbf{Source}\\ \hline \hline
    \multicolumn{3}{|c|}{\textbf{Hardware design related parameters}} \\ \hline
     $N_{g}^{2D}$ & User input (optional)&Input \\ \hline
    $N$ & $\geq 2$ &Input\\\hline
    $A_{die_{i}}$ & User input (optional)&Input/Eq.~(\ref{eq:3D_die_area_overall})   \\ \hline
    $N_{BEOL_i}$ & User input (optional)&Input/Eq.~(\ref{n_metal})\\ \hline
    
    \hline \hline
    
    \multicolumn{3}{|c|}{\textbf{Foundary related parameters}} \\ \hline
    Process & 3 $\sim$ 28 $nm$ & Input \\ \hline
    $A_{wafer_{die_i}}$ & 31,415.93 $\sim$ 159,043.13 $mm^2$ & Input \\ \hline
    $\gamma_{IO}^{micro\_3D/2.5D}$ & 0$\sim$1 &\cite{feng2022chiplet}\\ \hline
    $D_{TSV}$ &0.3 $\sim$25 $\mu m$ & \cite{kim2021micro} \\ \hline 
    $GPA_i, MPA_i$ & 0.1$\sim$0.5 $kg 
 \ CO_2/cm^2$&\cite{bardon2020dtco,gupta2022act}\\ \hline
   
    $EPA_i$ & 0.4$\sim$0.1 $ kWh/cm^2$&\cite{bardon2020dtco} \\ \hline
    $N_{fan}$ &1 $\sim$ 5 &\cite{stow2016cost}\\ \hline
    $p$ &0.6 $\sim$ 0.8 &\cite{stow2016cost}\\ \hline
    $\beta$ &450 $\sim$ 850 $M$&\cite{stow2016cost}\\ \hline
    $\lambda$ & 3 $\sim$ 28 $nm$&\cite{stow2016cost}\\ \hline
    $\omega$ & 3.6$\lambda$ &\cite{stow2016cost}\\
    \hline 
    \hline
    
    \multicolumn{3}{|c|}{\textbf{Bonding related parameters}} \\ \hline
    $CPA_{RDL}^{InFO_{2.5D}}$ & RDL characterization &\cite{bardon2020dtco}\cite{nagapurkar2022economic}\\ \hline
    $EPA_{D2W/W2W}^{micro/hybrid/C4}$ &0.9$\sim$2.75 $kwh/cm^2$&\cite{evg_bonding} \\ \hline
    $y^{micro/hybrid}_{W2W}$  &0$ \sim$ 1&\cite{evg_bonding}\cite{w2w}
    \\
    
    \hline \hline
    \multicolumn{3}{|c|}{\textbf{Substrate related parameters}} \\ 
    \hline
    $s_{RDL/EMIB/Si\_int}$ & $\geq 1$   &\cite{feng2022chiplet}\\ \hline
    $D_{gap}$ & 0.5$\sim$ 2 $mm$&\cite{feng2022chiplet} \\ \hline \hline

    \multicolumn{3}{|c|}{{\textbf{Packaging related parameters}}} \\ \hline
    $s_{package}^{3D/2.5D}$ & $\geq 1$ &\cite{nagapurkar2022economic}\cite{feng2022chiplet}\\ \hline
    $CPA_{packaging}$ & Package characterization&\cite{nagapurkar2022economic} \\ \hline \hline
    
    \multicolumn{3}{|c|}{\textbf{Carbon intensity}} \\ \hline
    $CI_{emb},CI_{use}$ & 30$\sim$ 700 $g\ CO_2\/kWh$ &\cite{gupta2022act}\\ \hline 
    
\end{tabular}
}

\vspace{-1.3em}
\end{threeparttable}

\label{tab:list_of_parameters}
\end{table}


\vspace{-0.5em}
\begin{align}
    DPW_{die_{i}/int}& = \frac{\pi \cdot \left(A_{wafer_{die_i}} / 2\right)^2}{A_{die_{i}/int}} - \frac{\pi \cdot A_{wafer_{die_i}}}{\sqrt{2 \cdot A_{die_{i}/int}}}
    \label{eq:DPW}
\end{align}

Similar to work~\cite{gupta2022act}, {\name} formulates the wafer carbon footprint ($C_{wafer_{die_i}}$) for $die_i$ as follows:

\begin{align}
C_{wafer_{die_i}} = & (CI_{emb} \cdot EPA_{i} + GPA_{i} + MPA_{i})\cdot A_{wafer_{die_i}}
\label{eq:wafer_carbon}
\end{align}

\noindent where $CI_{emb}$ is the carbon intensity of the fab's electrical grid (location), $EPA$/$GPA$/$MPA$ is fab energy/gas emissions/raw materials carbon foot print per unit 2D die area.

\noindent\textbf{Area Estimation:}
Additional areas for Through-Silicon Vias (TSVs) ($A^{3D}_{TSV_i}$) and interface I/O drivers ($A^{2.5D/Micro\_3D}_{IO_i}$) are required in 3D/2.5D ICs for die-to-die transmission. The total die area for $die_i$ includes gate area ($A_{gate_i}$), TSVs, and I/Os:

 \vspace{-0.7em}
\begin{align}
A_{die_{i}}^{3D/2.5D} &= A_{gate_i} + A^{3D}_{TSV_i} + A^{2.5D/Micro\_3D}_{IO_i}
\label{eq:3D_die_area_overall} 
\end{align}

\vspace{-0.3em}
Gate area ($A_{gate_i}$) is calculated from gate count ($N_{g_i}^{2D}$), feature size ($\lambda$), and scaling term ($\beta$):

\vspace{-0.7em}
\begin{align}
A_{gate_i} &=N_{g_i} ^{2D}\cdot \beta \cdot \lambda^2
\label{eq:2D_die_area}
\end{align}

$A^{3D}_{TSV_i}$ relates to the size of each TSV ($D_{TSV}$) and the TSV count ($X_{TSV_i}$). $D_{TSV}$ corresponds to each technode. Meanwhile, $X_{TSV_i}$ varies depending on the die stacking method (i.e., F2F/F2B). For F2B, the TSV count is calculated using Rent's rule as~\cite{stow2016cost}. F2F TSV count equals the IO number.



Considering the large size of micro-bumps and 2.5D connectors, additional I/O driver area ($A^{micro\_3D/2.5D}_{IO_i}$) is required, calculated using a ratio ($\gamma_{IO}^{micro\_3D/2.5D}$) of the gate area~\cite{feng2022chiplet}:

\vspace{-0.7em}
\begin{align}
    A^{micro\_3D/2.5D}_{IO_i}= \gamma_{IO}^{micro\_3D/2.5D} \cdot A_{gate_i}
\end{align}
 
\noindent\textbf{BEOL Configuration:} Reducing metal layers in the BEOL offers a more environmentally-friendly approach. The number of BEOL layers ($N_{BEOL_{i}}$) is estimated as follows~\cite{stow2016cost}:
\vspace{-0.5em}
\begin{align}
    & N_{BEOL_{i}} = \frac{N_{f a n} \cdot \omega \cdot N_{g_i} \cdot \bar{L_i}}{\eta \cdot A_{die_{i}}} 
\label{n_metal}
\end{align}

\subsubsection{$C_{bonding}^{3D/2.5D}$ Model}

3D/2.5D ICs involve wafer-to-wafer (W2W) or die-to-wafer (D2W) integration, stacking multiple dies or wafers~\cite{evg_bonding}. The bounding energy per unit area $EPA_{D2W/W2W}^{micro/hybrid/C_4}$ and yield $Y_{bonding_i}^{micro/hybrid/C_4}$ depends on the choice of D2W or W2W, and on the bonding method (C4-bump, micro-bumping, or hybrid bonding).

\begin{align}
C_{bonding}^{3D/2.5D} = \sum_{i=1}^{N-1}\frac{ CI_{emb} \cdot EPA_{D2W/W2W}^{micro/hybrid/C4} \cdot A_{die_{i}}}{Y_{bonding_i}^{micro/hybrid/C4}} 
\label{eq:carbon_embodied_3D_bonding}
\end{align}
\vspace{-0.7em}

\subsubsection{$C_{packaging}^{3D/2.5D}$ Model}
In {\name}, the packaging carbon footprint is estimated using packaging carbon emissions per area ($CPA_{packaging}$):

\begin{align}
C_{packaging}^{3D/2.5D} &= CPA_{packaging} \cdot A_{package}^{3D/2.5D}
\label{eq:carbon_embodied_3D/2.5D_packaging}
\end{align}

The package area $A_{package}^{3D/2.5D}$ is calculated using a linear empirical equation from~\cite{feng2022chiplet}, applying a scaling factor $s_{package}^{3D/2.5D}$, which is based on the largest die area in 3D ICs and the total die area for 2.5D ICs.

\subsubsection{$C_{int}^{2.5D}$ Model}
The carbon footprint for RDL, EMIB, and silicon interposer substrates is modeled similarly to die carbon footprint. The interposer area calculation differs for each integration method:

\vspace{-0.7em}
\begin{align}
   & A_{Si\_int} = s_{Si\_int} \cdot \sum_{i=1}^N A_{die_{i}}  \\
   & A_{RDL/EMIB} = s_{RDL/EMIB} \cdot D_{gap}  \cdot\sum_{i=1}^N l_{adjacent_i} 
\end{align}

\noindent where $s_{Si\_int}$ and $s_{RDL/EMIB}$ are a scaling factors, $l_{adjacent_i}$ is the total length of adjacent sides for all dies, and $D_{gap}$ is the gap between two adjacent dies.

\subsubsection{Yield Model}
\label{sec: yield}
We estimate yields for different process technologies using data from \cite{nagapurkar2022economic} for bonding and packaging yields, and a yield distribution model from \cite{feng2022chiplet} for die and substrate yields.

\vspace{-0.7em}
\begin{align}
y_{die_{i}}=(1+\frac{A_{d i e_i}(d, p) \times D_0}{\alpha})^{-\alpha}
\end{align}

\noindent where $D_0$ is the defect density and $\alpha$ is a parameter determined by process complexity which both given by \cite{feng2022chiplet}. 

We also consider the impact of each process on the overall yield. Denoting individual process yield as $y$ and overall yield as $Y$, the yield of a process like W2W bonding ($Y_{bonding_i}$) is affected by its own yield ($y_{b}$) and the die yield ($y_{die}$), as defective dies cannot be separated before bonding. The results for different $Y$ values are listed in Tab.~\ref{tab: yield}.

\begin{table}[t]
\centering
\footnotesize
\caption{Stacking Yields}
\vspace{-1.2em}
\resizebox{1\linewidth}{!}{ 
\def\arraystretch{1.5}\tabcolsep 2pt
\def\thefootnote{a}\footnotesize
\begin{tabular}{|c|c|c|}
\hline
\textbf{3D Type} & \textbf{$Y_{die_{i}}^{micro/hybrid}$} & \textbf{$Y_{bonding_i}^{micro/hybrid}$} \\
\hline
D2W & $y_{die_{i}} \cdot (y^{micro/hybrid}_{D2W}) ^{N-i
}$ & $(y^{micro/hybrid}_{D2W}) ^{N-i}$ \\
\hline
W2W & $\prod_{j=1}^N y_{die_j} \cdot( y^{micro/hybrid}_{W2W})^{N-1} $& $\prod_{j=1}^N y_{die_j} \cdot( y^{micro/hybrid}_{W2W})^{N-1} $  \\
\hline
\end{tabular}}
\centering
\resizebox{0.9\linewidth}{!}{ 
\def\arraystretch{1.5}\tabcolsep 2pt
\def\thefootnote{a}\footnotesize
\begin{tabular}{|c|c|c|c|}
\hline
\textbf{2.5D Type} & \textbf{$Y_{die_{i}}^{2.5D}$} & \textbf{$Y_{substrate}^{2.5D}$} & \textbf{$Y_{bonding_i}^{2.5D}$} \\
\hline
Chip-first & $y_{die_{i}} \cdot y_{substrate}^{2.5D}$ & $y_{substrate}^{2.5D}$ &1\\
\hline
Chip-last & $y_{die_{i}} \cdot \prod_{j=1}^{N}y^{2.5D}_{bonding_j}$ & $y_{substrate}^{2.5D} \cdot \prod_{j=1}^{N}y^{2.5D}_{bonding_j}$ & $\prod_{j=1}^N y^{2.5D}_{bonding_j}$\\
\hline

\end{tabular}
}
\vspace{-1.2em}
\label{tab: yield}
\end{table}

\subsection{{\name}: Operational Carbon Emission}

We focus on the fixed-throughput approach, widely adopted in applications like autonomous vehicles (AVs)~\cite{AV}. Given the varied energy consumption patterns that different power benchmarks can produce, the operational carbon footprint ($C_{operational}$) of ICs is determined by the diverse run-time ($T_{app_k}$), the use carbon intensity ($CI_{use}$), and power consumption ($P_{app_k}$) of each application.

\vspace{-0.7em}
\begin{align}
C_{opreational} =\sum_{k} CI_{use} \cdot P_{app_k} \cdot T_{app_k}
\label{eq:carbon_operational}
\end{align}



\vspace{-0.7em}
The power ($P_{app_k}$) is calculated as:

\begin{align} 
\label{eq:power}
   & P_{app_k}= \sum_{i=1}^{N}(\frac{Th_{app_k}}{Eff_{die_{i}}} + P_{IO_i}) \nonumber \\ 
   & P_{IO_i}=P_{per\_pitch_{i}} \cdot N_{pitch_{i}} \nonumber\\
   & N_{pitch_{i}}=L_{edge_{i}} \cdot D_{pitch_{i}} \cdot N_{BEOL_{i}}  
\end{align}

\noindent where $Th$ is the fixed throughput, $Eff_{die_{i}}$ is the energy efficiency of $die_i$, and $P_{IO_i}$ is each die's interface I/O driver power. In the absence of specific input for $Eff_{die
_i}$, we utilize surveyed parameters (e.g., as in \cite{kim2021micro}) to estimate $Eff_{die_{i}}$. For 2.5D ICs and Micro-bumping 3D ICs, the I/O power ($P_{IO_i}$) should be included. We presume $P_{IO_i}$, using the energy cost per pitch ($P_{per\_{pitch}_{i}}$), the length of $die_i$'s edge ($L_{edge_{i}}$), the surveyed pitch density ($D_{pitch_{i}}$), and the number of BEOL layers ($N_{BEOL_{i}}$).


\subsection{\name: Bandwidth Constraint}
A key constraint for 3D/2.5D ICs is ensuring sufficient I/O interface bandwidth for applications. We assume that 3D ICs' I/O bandwidth matches the on-chip bandwidth of their 2D counterparts~\cite{MCMGPU}. 
The 2.5D ICs' I/O bandwidth is:
\label{sec: constrains}

\begin{equation}
    BW_{die_{i}}= N_{I/O_{i}}\cdot BW_{per\_I/O_{i}}
\end{equation}

With deep neural networks being the primary workload for AVs~\cite{AV}, we establish the bandwidth constraint that the performance (i.e., throughput) degradation of 2.5D ICs exceeds 20\% if the I/O bandwidth is reduced by half than the 2D on-chip bandwidth~\cite{MCMGPU}. 
Based on this, if the 2.5D ICs’ interface causes performance to fall below the throughput requirement, we categorize these instances as ``invalid''.

\begin{figure}[t]
  \centering
  \includegraphics[width=1\linewidth]{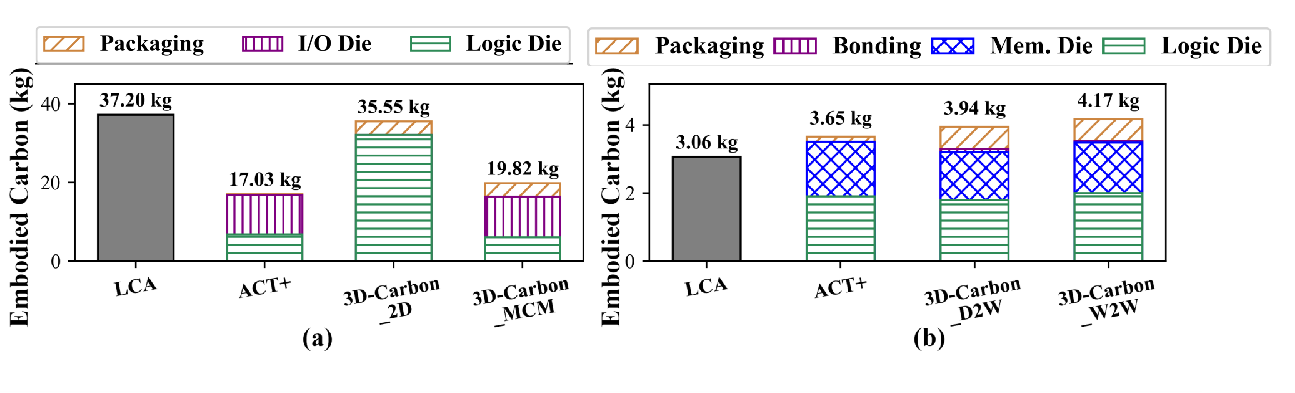}
\vspace{-3em}
  \caption{Validation of {\name} against (a) 2.5D EPYC 7452~\cite{EPYC} and (b) 3D Lakefield~\cite{lakefield}.}
\vspace{-0.7em}
\label{fig:EXP0}
\end{figure}

%% file: Sections/validation.tex
\section{Validation of {\name}}
We validate {\name} by comparing its predicted embodied carbon emissions with those obtained from Life Cycle Assessment (LCA) reports~\cite{Gabi} and the latest ACT+ tool~\cite{gupta2022act}.

\subsection{Validation against One 2.5D IC: EPYC 7452}
We validate our {\name} model against one MCM 2.5D IC, EPYC 7452~\cite{EPYC}. Inputs for both {\name} and ACT+ are based on the EPYC 7452's specifications: 7nm technology for four CPU dies and 14nm for one I/O die.

As shown in Fig.~\ref{fig:EXP0}(a), the LCA~\cite{Gabi}, designed for 2D monolithic ICs, reports higher embodied emissions than {\name} and ACT+. When we adjust {\name} for a 2D IC, the discrepancy in embodied emissions between LCA and {\name} is about 4.4\%. Unlike ACT+, our model includes manufacturing complexity: it considers BEOL configurations, adjusting carbon footprint for CPU dies with fewer BEOL layers, and estimates packaging carbon emissions based on actual packaging area, resulting in higher packaging carbon emission (3.47 kg) compared to ACT+'s fixed 0.15 kg carbon.

\subsection{Validation against One 3D IC: Lakefield}

We validate our model using Intel's Lakefield 3D IC~\cite{lakefield}, which features heterogeneous integration with a 7nm top logic die and a 14nm base memory die, and show the results in Fig.~\ref{fig:EXP0}(b). Both {\name} and ACT+ use the Lakefield's technology specifications.

We reference the GaBi LCA database~\cite{Gabi} for LCA validation baseline. Since GaBi doesn't cover the 7 nm process, it assume 14nm for both dies, leading to an underestimation than {\name} and ACT+ as the 7nm node is more complex and carbon-intensive. Compared to ACT+, our model accounts for manufacturing complexities and differences between D2W and W2W stacking methods. D2W, involving advanced bonding technology, results in lower yield for the bonding process but allows pre-stacking die availability checks, leading to higher individual die yields ($Y_{die}$). Specifically, the logic die yield in D2W is 89.3\%, the memory die is 88.4\%, whereas in W2W, both dies have a yield of 79.7\%.

\section{Case study: Sustainable Decision-Making for NVIDIA Autonomous Vehicle GPUs}  

We conduct sustainable decision-making case studies using \name\ on the NVIDIA GPU DRIVE series for AVs as detailed in Tab.~\ref{tab:gpu}. These compare carbon emissions of original 2D designs with hypothetical 3D/2.5D designs. The hypothetical designs involve two die division approaches: homogeneous (splitting the 2D IC into two similar dies) and heterogeneous (isolating the memory and IOs from the main logic die and implementing them separately in an older 28nm node). For 3D ICs, we consider F2F with D2W stacking.

\begin{table}[t]
    \caption{NVIDIA GPU DRIVE series specifications~\cite{nvidia}.}
    \vspace{-1.2em}
    \centering 
    \resizebox{0.95\linewidth}{!}{   
    \begin{tabular}{|c|c|c|c|c|}
       \hline
        &PX 2&XAVIER&ORIN&THOR\\ \hline
        Technology node (nm)  &16& 12& 7&5\\ \hline
        Gate count (Billion) &15.3&21&17&77 \\ \hline
        Energy efficiency (TOPS/W)&0.75&1&2.74&12.5\\ \hline
        Announced (year)&2016&2017&2019&2022\\ \hline
    \end{tabular}}
    \label{tab:gpu}
    \vspace{-1.2em}
\end{table}

\begin{figure}[t]
  \centering
  \includegraphics[width=1\linewidth]{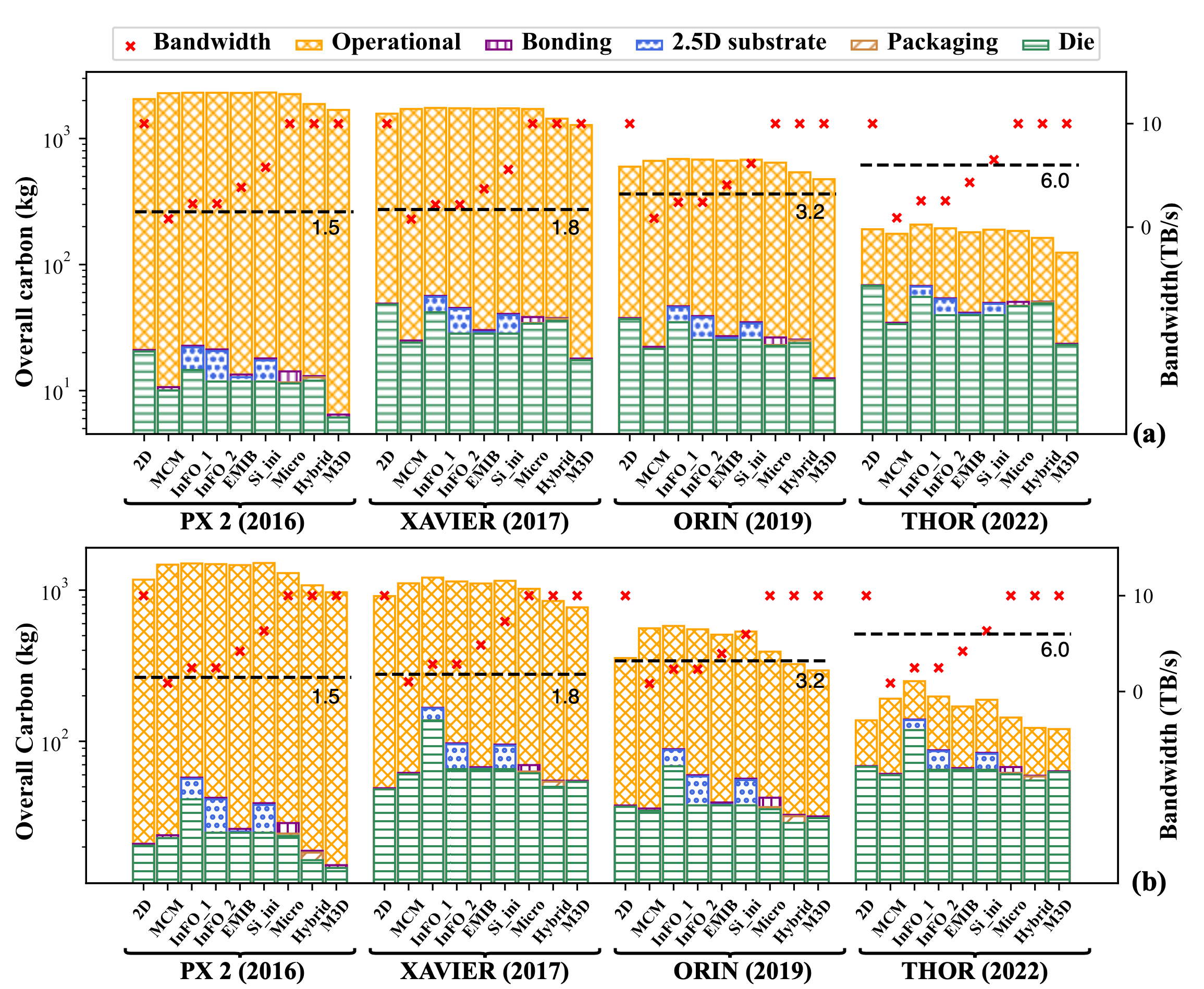}
  \vspace{-2.5em}
  \caption{Overall carbon emissions of NVIDIA DRIVE series: 2-die 3D/2.5D ICs with the (a) homogeneous and (b) heterogeneous approaches~\cite{MCMGPU}. Note that InFO\_1/InFO\_2 represent chip-first/chip-last approaches, respectively.}
  \vspace{-1.em}
  \label{fig:vehicles}
\end{figure}

\subsection{Overall Carbon Footprint}

\vspace{-0.4em}
Fig.~\ref{fig:vehicles}(a) and (b) show NVIDIA DRIVE series' carbon emissions for homogeneous and heterogeneous 3D/2.5D approaches, respectively. The black line indicates the required I/O bandwidth, and the red ``x'' marks achieved bandwidth. 
InFO and silicon-interposer 2.5D ICs increase embodied carbons due to large substrate areas and low substrate yields. Other 3D/2.5D designs constantly reduce/maintain the embodied carbons, particularly in the homogeneous approach (see Fig.~\ref{fig:vehicles}(a)), while the heterogeneous approach (see Fig.~\ref{fig:vehicles}(b)) introduces lesser saving due to smaller memory die areas and limited benefits from the older technology. 
With the exponential growth of energy efficiency over time, the operational carbon emissions decrease, as detailed in Tab.~\ref{tab:gpu}.
Operational carbon emissions are higher for 2.5D ICs than 2D/3D ICs, due to the performance degradation (i.e., throughput) from {\name}'s bandwidth constraint (see Sec.~\ref{sec: constrains}). For THOR, none of the four 2.5D ICs meet the necessary bandwidth, rendering them ``invalid''.

\subsection{Sustainable Decision-Making}
We chose the five valid 3D/2.5D ICs with the homogeneous approach for NVIDIA DRIVE ORIN. Tab.~\ref{tab:nvidia} presents the embodied carbon savings, overall carbon savings, and metrics for choosing and replacing relative to the original 2D IC. These 3D/2.5D ICs can save up to 65.53\% embodied carbon emission and up to 41.03\% overall carbon emission.  
Given the average 10-year lifetime of AV devices, the EMIB 2.5D IC and all three types of 3D ICs can save carbon emissions compared to the 2D IC, as the 10-year lifetime falls within their choosing metric ($T_c$) ranges. 
For the decision-making of replacing the 2D IC with 3D/2.5D ICs, we advise against replacing the original 2D IC due to the significant embodied carbon emissions in the new 3D/2.5D ICs, which cannot be compensated by operational carbon savings over their lifetime (see their replacing metric ($T_r$) ranges).

\begin{table}[t]
    \centering
    \caption{Case studies for choosing/replacing the NVIDIA DRIVE ORIN 2D IC with the 3D/2.5D ICs.}
    \vspace{-1em}
    \resizebox{1\linewidth}{!}{
    \begin{tabular}{|l|c|c|c|c|c|}
    \hline
    3D/2.5D ICs &EMIB&Si\_int&Micro&Hybrid&M3D\\
    \hline
      Embodied carbon save ratio&23.69\% & -9.59\% &25.88\% &35.64\% &65.53\%  \\ \hline
     
Overall carbon save ratio&6.5\% & -9.86\% &7.63\% &21.71\% &41.03\%  \\ \hline
Choosing metric $T_c$ (years)&$<$22 &$\infty$& $<$25 & $>$0 & $>$0 \\ \hline
Replacing metric $T_r$ (years)&$\infty$ &$\infty$ &$\infty$ &$>$ 75 &$>$ 19 \\ \hline
    \end{tabular}}
    \label{tab:nvidia}
\vspace{-1.em}
\end{table}

%% file: Sections/conclusion.tex
\section{Conclusion}
This work introduces {\name}, an analytical carbon modeling tool designed to understand the carbon emissions of commercial-grade 3D/2.5D ICs at the early design stage. 
Addressing the need for such tools amidst the growing use of advanced integration technologies, {\name} aims to pave the way for future developments in environmentally sustainable 3D and 2.5D ICs.



\section*{Acknowledgement}
This work was supported in part by CoCoSys, one of the seven centers in JUMP 2.0, a Semiconductor Research Corporation (SRC) program sponsored by DARPA.